\documentclass[letterpaper,11pt]{article}
\pdfoutput=1
\usepackage{jcappub}
\usepackage{ifpdf}
\usepackage{graphicx}
\usepackage{amsmath}
\usepackage{amssymb}

\notoc
\title{The Consistency of Fermi-LAT
  Observations of the Galactic Center with a Millisecond Pulsar
  Population in the Central Stellar Cluster} 

\author{Kevork N.\ Abazajian}
\affiliation{Maryland Center for Fundamental Physics \& Joint
  Space-Science Institute, Department of Physics, University of
  Maryland, College Park, Maryland 20742 USA}
\emailAdd{kev@umd.edu}

\abstract{
  I show that the spectrum and morphology of a recent Fermi-LAT
  observation of the Galaxy center are consistent with a millisecond
  pulsar population in the nuclear Central stellar cluster of the
  Milky Way.  The Galaxy Center gamma-ray spectrum is consistent with
  the spectrum of four of eight globular clusters that have been
  detected in the gamma-ray.  A dark matter annihilation
  interpretation cannot be ruled out, though no unique features exist
  that would require this conclusion.
}

\arxivnumber{arXiv:1011.4275}
\subheader{UMD-PP-10-019}


\begin{document}

\maketitle
\flushbottom

\section{Introduction } The ability of gamma-ray
observatories to provide a window on dark matter annihilation signals
has been known for some time, (e.g.~\cite{Jungman:1995df}).  
The recent
manuscript by Hooper \& Goodenough~\cite{Hooper:2010mq} performs a
detailed analysis to extract the signal in the gamma-ray of the
dynamical center of our Milky Way Galaxy as observed by the Large Area
Telescope (LAT) aboard the Fermi Gamma-Ray Space Telescope.  That work
interpreted the morphology and spectral feature of the signal within
the inner 1.25$^\circ$ (175 parsec radius) from the Galactic dynamic
center to be inconsistent with any known astrophysical sources, and
further claimed for it to be accounted for by the presence of
annihilating light ($\sim$8 GeV) dark matter.  In this short note, I
show that the signal as seen by Hooper \& Goodenough (hereafter HG) is
from the Galactic Central stellar cluster and is consistent with prior
observations of gamma-ray emission from massive stellar globular
clusters with a population of millisecond pulsars (MSP)
\cite{Abdo:2010bb}.

\begin{figure}[t]
\includegraphics[width=6.truein]{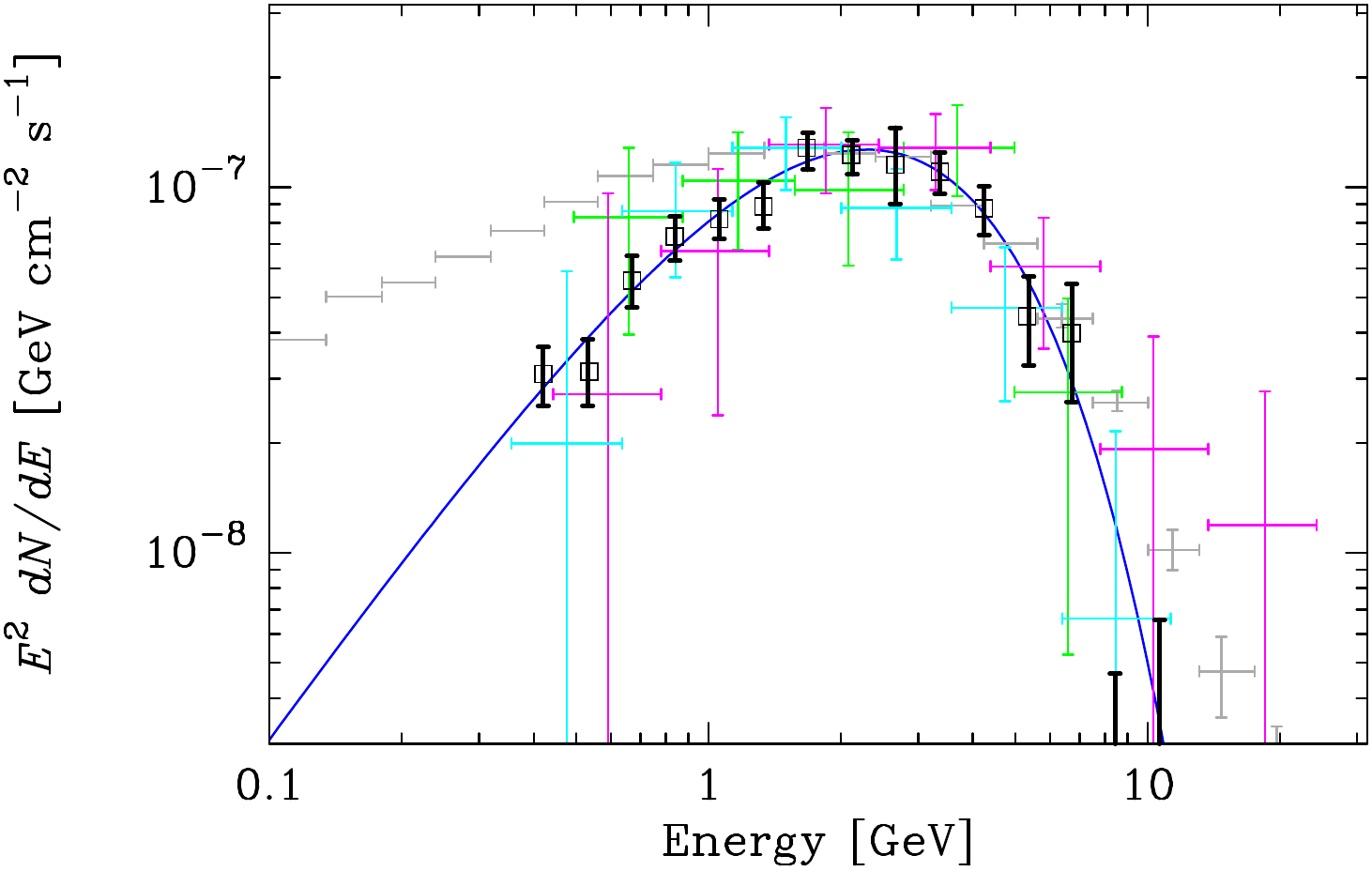}
\caption{Shown is the spectrum $E^2 dN/dE$ of the HG source (thick
  black points with errors), and its best fit power law with
  exponential cutoff curve (blue).  
  The HG spectrum power-law index is consistent with the
  globular cluster gamma ray emission of Omega Cen (green), NGC
  6388 (cyan) and
  M 28 (magenta) \cite{Abdo:2010bb}.  The spectrum of the Geminga
  pulsar (light grey) \cite{Abdo:2010wp} has a nearly identical peak
  energy and
  exponential cutoff and a softer power-law index than the HG source.  The
  spectra of Omega Cen and M 28 are shifted by +0.14 dex and +0.34
  dex
  in energy, respectively, to illustrate the consistency in power-law
  index.  NGC 6388 is not shifted in energy.  The HG source is
  consistent, within errors, with the shape of the intrinsic spectrum of 
  all the plotted globular
  cluster sources.  All sources are normalized to the HG peak
  flux. 
  \label{spectrum}}
\end{figure}

\section{Spectrum}
In general, pulsar gamma-ray spectra have a flux peak at $\sim 2-3$
GeV and a spectrum $dN/dE \propto E^{-\Gamma} \exp(E/E_{\rm c})$ with
a power law with index $\Gamma$ in the range 0.5-2.5 and exponential
cutoff energy $E_{\rm c}$ at 1-5 GeV, a range consistent with the HG
source \cite{Abdo:2009ax}.  Note that HG compare to the {\it average}
millisecond pulsar spectrum and find it is not consistent, and then
reach the conclusion that no astrophysical source can produce the HG
spectrum. However, {\it this is not appropriate}, but rather
consistency with {\it any} set of pulsars or stellar cluster population
of pulsars must be ruled out before claiming inconsistency with
astrophysical sources.

The spectrum of the HG source can be well fit by a pulsar-like
power law plus exponential cutoff spectrum.  The spectrum below 7 GeV
is well fit by $dN/dE \propto E^{-\Gamma} \exp(E/E_{\rm c})$, with $\Gamma =
0.29 \pm 0.12\rm\ GeV$ and $E_{\rm c} = 1.34\pm 0.12\rm\ GeV$ and a reduced
$\chi^2/{\rm DOF} = 0.5$. The spectrum and best fit curve is shown in
Fig.~(\ref{spectrum}).

Several pulsars in the First Fermi-LAT Catalog of Gamma-ray
Pulsars~\cite{Abdo:2009ax}, including J1958+2846, J2032+4127 and
J2043+2740, have a power-law index and exponential cutoff consistent
with the HG source.  The Geminga pulsar has a detailed published
gamma-ray spectrum which has a similar spectral peak, $2-3\rm\ GeV$
and cutoff as the HG source~\cite{Abdo:2010wp}.  The Geminga spectrum
is shown for comparison in Fig.~(\ref{spectrum}). However, Geminga has
a shallower power-law spectral index $\Gamma = 1.08\pm 0.02$.

However, it is not expected that the extreme density of the Galactic
Central stellar cluster harbors only a single pulsar.  The Central
stellar cluster (nuclear bulge) of the Milky Way is coincident with the
observational region of HG central spectral signal, and extends across
the full $\sim 1 ^\circ$ from the dynamical center (or $\sim$300 pc
across) of that spectral analysis region.  The spectrum of the HG
Central stellar cluster region ($\Gamma = 0.29 \pm 0.12$, $E_{\rm c} =
1.34\pm 0.12$) is consistent with four of the eight globular clusters
that have been observed by Fermi-LAT and seen to have gamma-ray
emission \cite{Abdo:2010bb}:
\begin{eqnarray}
\text{Omega Cen:} &\quad&  \Gamma =
0.7^{+0.7+0.4}_{-0.6-0.4}, E_{\rm c} = 1.2^{+0.7+0.2}_{-0.4-0.2},\nonumber \\
\text{NGC 6388:} &\quad& \Gamma = 1.1^{+0.7+0.8}_{-0.5-0.8}, E_{\rm c} =
  1.8^{+1.2+1.8}_{-0.7-0.6},\nonumber \\
\text{M 28:} &\quad& \Gamma = 1.1^{+0.7+0.6}_{-0.5-0.7}, E_{\rm c} =
  1.0^{+0.6+0.4}_{-0.3-0.2},\nonumber  \\
\text{NGC 6652:} &\quad& \Gamma = 1.0^{+0.6+0.3}_{-0.5-0.3}, E_{\rm c} =
  1.8^{+1.2+0.4}_{-0.6-0.3}.
\label{fitsglobclust}
\end{eqnarray}
Three of these spectra are shown in Fig.~(\ref{spectrum}).  Note that
a direct quantitative comparison of the fit to the HG source spectrum
presented above with the globular cluster fits,
Eqs.~(\ref{fitsglobclust}), cannot be rigorously performed since the
sets of fits use different methods.  The fit performed in this
analysis is to the points and errors presented in HG, while the fits
in Eqs.~(\ref{fitsglobclust}) from Ref.~\cite{Abdo:2010bb} are
performed with the Fermi-LAT analysis tool \texttt{gtlike}, which bins
the spectrum differently and more accurately incorporates the photon
statistics.  The fit between HG and the globular clusters are
obviously qualitatively consistent (Fig.~\ref{spectrum}), but a
detailed statistical comparison can only be established with
\texttt{gtlike} analyses of all regions. Most importantly,
over-subtraction or under-subtraction of the diffuse component of an
observation can lead to a systematic uncertainty that must also be
included in a full error analysis of the Galaxy Central stellar
cluster spectrum \cite{Ferrara:2010}.  This was not performed in the
HG analysis.  Over-subtraction of the diffuse component may lead to
a high estimate of the intrinsic spectral index $\Gamma$.

\section{Morphology and Luminosity} 
The Central stellar cluster of the Milky Way has a stellar mass of
approximately $4\times 10^9 M_\odot$~\cite{Mezger:1996} and harbors
2357 X-ray point sources within the central $17' \times 17'$
alone~\cite{Muno:2003}.  In comparison, Omega Cen is the most massive
stellar globular cluster in our Galaxy's halo with mass $5 \times 10^6
M_\odot$ \cite{Meylan:1995}, and has 45-70 X-ray sources associated
with it~\cite{Haggard:2009}.

The peak energy flux in gamma rays of the Central stellar cluster
associated with the HG source is approximately 200 times that of Omega
Cen, when corrected for distance.  Given the three order of magnitude
greater mass scale of the Central Galactic stellar cluster and its
relative high-energy activity to that of the globular clusters, the
factor of $\sim$200 enhancement of flux is not unreasonable.

In the same regard, the gamma ray profile of a globular cluster like
Omega Cen only extends to $7.5'$, while that observed in the Central
stellar cluster by HG is approximately $30'$, which can also be
accounted for by the 3 order of magnitude larger mass scale of the
Central stellar cluster.

\section{Conclusions}
The gamma-ray flux spectrum and profile observed by HG towards the
Central stellar cluster of the Milky Way is consistent with other
stellar cluster populations in observed globular clusters. The
spectral index, exponential cutoff and peak flux energy is consistent
with four of the eight detected globular clusters in the gamma-ray by
Fermi-LAT.  Therefore, there exists is no feature of the HG spectral
source that necessarily requires a dark matter annihilation
interpretation.

Of the detected gamma-ray observed globular clusters, only 5 of 8
harbor detected MSPs~\cite{Abdo:2010bb}.  The gamma-ray emission from
stellar clusters such as the Central stellar cluster as well as
globular clusters is expected to be dominated by MSP emission due to
enhanced binary formation in these systems, where MSPs are spun up by
accretion from their binary partners.  Therefore, it is likely that
the detection of the peaked spectrum towards the Galactic Central
stellar cluster by Hooper \& Goodenough is that of a population of
millisecond pulsars bound within this massive stellar cluster.
Obscuration, crowding, and absorption towards the central region of
the Milky Way makes direct detection of the MSPs difficult, though
this identification of the spectral signature of stellar cluster-like
MSPs in the Galactic Center may motivate deeper searches.

\noindent {\it Acknowledgments --- } I would like to thank Prateek
Agrawal, Steve Blanchet, Z.\ Chacko, J.\ Pat Harding, Manoj
Kaplinghat, J\"urgen Kn\"odlseder, Julie McEnery, and Paul Ray for
useful discussions.  In particular, I would like to thank Elizabeth
Ferrara for discussions and detailed comments on the manuscript, as
well as her and Chris Shrader for organizing a GSFC Fermi-LAT Science
Workshop on November 16, 2010 which led to this analysis.  K.A. is
supported by NSF Grant 07-57966 and NSF CAREER Award 09-55415.

\bibliography{master}

\providecommand{\href}[2]{#2}\begingroup\raggedright\begin{thebibliography}{10}

\bibitem{Jungman:1995df}
G.~Jungman, M.~Kamionkowski, and K.~Griest, {\it {Supersymmetric dark matter}},
   {\em Phys. Rept.} {\bf 267} (1996) 195--373,
  [\href{http://xxx.lanl.gov/abs/hep-ph/9506380}{{\tt hep-ph/9506380}}].

\bibitem{Hooper:2010mq}
D.~Hooper and L.~Goodenough, {\it {Dark Matter Annihilation in The Galactic
  Center As Seen by the Fermi Gamma Ray Space Telescope}},
  \href{http://xxx.lanl.gov/abs/1010.2752}{{\tt arXiv:1010.2752}}.

\bibitem{Abdo:2010bb}
{\bf Fermi-LAT} Collaboration, A.~Abdo {\em et.~al.}, {\it {A population of
  gamma-ray emitting globular clusters seen with the Fermi Large Area
  Telescope}},  \href{http://xxx.lanl.gov/abs/1003.3588}{{\tt
  arXiv:1003.3588}}.

\bibitem{Abdo:2010wp}
{\bf Fermi-LAT} Collaboration, A.~Abdo {\em et.~al.}, {\it {Fermi LAT
  observations of the Geminga pulsar}},
  \href{http://xxx.lanl.gov/abs/1007.1142}{{\tt arXiv:1007.1142}}.

\bibitem{Abdo:2009ax}
{\bf Fermi-LAT} Collaboration, A.~Abdo {\em et.~al.}, {\it {The First Fermi
  Large Area Telescope Catalog of Gamma-ray Pulsars}},  {\em
  Astrophys.J.Suppl.} {\bf 187} (2010) 460--494,
  [\href{http://xxx.lanl.gov/abs/0910.1608}{{\tt arXiv:0910.1608}}].

\bibitem{Ferrara:2010}
E.~Ferrara {\em private communication} (2010).

\bibitem{Mezger:1996}
P.~G. {Mezger}, W.~J. {Duschl}, and R.~{Zylka}, {\it {The Galactic Center: a
  laboratory for AGN?}},  {\em \aapr} {\bf 7} (1996) 289--388.

\bibitem{Muno:2003}
M.~P. {Muno}, F.~K. {Baganoff}, M.~W. {Bautz}, W.~N. {Brandt}, P.~S. {Broos},
  E.~D. {Feigelson}, G.~P. {Garmire}, M.~R. {Morris}, G.~R. {Ricker}, and L.~K.
  {Townsley}, {\it {A Deep Chandra Catalog of X-Ray Point Sources toward the
  Galactic Center}},  {\em \apj} {\bf 589} (May, 2003) 225--241,
  [\href{http://xxx.lanl.gov/abs/astro-ph/}{{\tt astro-ph/}}].

\bibitem{Meylan:1995}
G.~{Meylan}, M.~{Mayor}, A.~{Duquennoy}, and P.~{Dubath}, {\it {Central
  velocity dispersion in the globular cluster {$\omega$} Centauri.}},  {\em
  \aap} {\bf 303} (Nov., 1995) 761--+.

\bibitem{Haggard:2009}
D.~{Haggard}, A.~M. {Cool}, and M.~B. {Davies}, {\it {A Chandra Study of the
  Galactic Globular Cluster Omega Centauri}},  {\em \apj} {\bf 697} (May, 2009)
  224--236, [\href{http://xxx.lanl.gov/abs/0902.2397}{{\tt arXiv:0902.2397}}].

\end{thebibliography}\endgroup

\bibliographystyle{jhep}

\end{document}